\def\etc{{\it etc.}}

\def\ie{{\it i.e.}}
\def\vs{{\it vs.}}

\def\~{{$\tilde{\phantom{a}}$}}

\documentclass [12pt] {article}
\usepackage{epsfig}
\usepackage{color}

\renewcommand{\arraystretch}{1.5}

\def\thebibliography#1{\section{References}\markboth
 {REFERENCES}{REFERENCES}\list
 {[\arabic{enumi}]}{\settowidth\labelwidth{[#1]}\leftmargin\labelwidth
 \advance\leftmargin\labelsep
 \usecounter{enumi}}
 \def\newblock{\hskip .11em plus .33em minus -.07em}
 \sloppy
 \sfcode`\.=1000\relax}
\def\upcite#1{\raise6pt\hbox{\scriptsize
\cite{#1}}}
  \def\lsim{\mathrel {\vcenter {\baselineskip 0pt \kern 0pt
    \hbox{$<$} \kern 0pt \hbox{$\sim$} }}}
    \def\gsim{\mathrel {\vcenter {\baselineskip 0pt \kern 0pt
    \hbox{$>$} \kern 0pt \hbox{$\sim$} }}}

\def\hline{\noalign{\hrule \vskip2pt}}

\def\|{\ifmmode\Vert\else \char`\|\fi}
\ifx\oldzeta\undefined                          % hasn't been done yet, so 
  \let\oldzeta=\zeta                            % save old definiton
  \def\zzeta{{\raise 2pt\hbox{$\oldzeta$}}}     % make new definition
  \let\zeta=\zzeta                              % and attatch it
\fi

\ifx\oldchi\undefined                           % hasn't been done yet, so 
  \let\oldchi=\chi                              % save old definiton
  \def\cchi{{\raise 2pt\hbox{$\oldchi$}}}       % make new definition
  \let\chi=\cchi                                % and attatch it
\fi

\def\frac#1#2{{#1 \over #2}}

\def\half{\ifinner {\scriptstyle {1 \over 2}}
   \else {1 \over 2} \fi}

\def\abs#1{\left\vert#1\right\vert}	% \abs{stuff} gives |stuff|
\def\simge{\mathrel{%
   \rlap{\raise 0.511ex \hbox{$>$}}{\lower 0.511ex \hbox{$\sim$}}}}
\def\simle{\mathrel{
   \rlap{\raise 0.511ex \hbox{$<$}}{\lower 0.511ex \hbox{$\sim$}}}}

\def\buildchar#1#2#3{{\null\!                   % \null, cancel space
   \mathop#1\limits^{#2}_{#3}                   % #1, #2 above, #3 below
   \!\null}}                                    % cancel space, \null
\def\overcirc#1{\buildchar{#1}{\circ}{}}

\def\slashchar#1{\setbox0=\hbox{$#1$}           % set a box for #1 
   \dimen0=\wd0                                 % and get its size
   \setbox1=\hbox{/} \dimen1=\wd1               % get size of /
   \ifdim\dimen0>\dimen1                        % #1 is bigger
      \rlap{\hbox to \dimen0{\hfil/\hfil}}      % so center / in box
      #1                                        % and print #1
   \else                                        % / is bigger
      \rlap{\hbox to \dimen1{\hfil$#1$\hfil}}   % so center #1
      /                                         % and print /
   \fi}                                         %

\def\subrightarrow#1{%                          % #1 under arrow
  \setbox0=\hbox{%                              % set a box
    $\displaystyle\mathop{}%                    % no mathop
    \limits_{#1}$}%                             % just limits
  \dimen0=\wd0%                                 % get width
  \advance \dimen0 by .5em%                     % add a bit
  \mathrel{%                                    % space like =
    \mathop{\hbox to \dimen0{\rightarrowfill}}% % arrow to width
       \limits_{#1}}}                           % text below

\def\overlay#1#2{\ifmmode%
\setbox0=\hbox{$#1$}%
\setbox1=\hbox to\wd0{\hss$#2$\hss}\else%
\setbox0=\hbox{#1}%
\setbox1=\hbox to\wd0{\hss#2\hss}\fi%
#1\hskip-\wd0\box1 }

\def\pmb#1{\leavevmode\setbox0=\hbox{#1}%
\kern-.02em\copy0\kern-\wd0
\kern.04em\copy0\kern-\wd0
\kern-.02em\raise.04em\box0 }

\def\vereq#1#2{\lower3pt\vbox{\baselineskip1.5pt \lineskip1.5pt
\ialign{$\m@th#1\hfill##\hfil$\crcr#2\crcr\sim\crcr}}}

\def\tensor#1{\protect\@ontopof{#1}{\leftrightarrow}{1.15}\mathord{\box2}}
\def\overstar#1{\protect\@ontopof{#1}{\ast}{1.15}\mathord{\box2}}
\def\overdots#1{\protect\@ontopof{#1}{\cdots}{1.0}\mathord{\box2}}
\def\overcirc#1{\protect\@ontopof{#1}{\circ}{1.2}\mathord{\box2}}
\def\loarrow#1{\protect\@ontopof{#1}{\leftarrow}{1.15}\mathord{\box2}}
\def\roarrow#1{\protect\@ontopof{#1}{\rightarrow}{1.15}\mathord{\box2}}

\def\@ontopof#1#2#3{%
{\mathchoice
{\@@ontopof{#1}{#2}{#3}\displaystyle\scriptstyle}%
{\@@ontopof{#1}{#2}{#3}\textstyle\scriptstyle}%
{\@@ontopof{#1}{#2}{#3}\scriptstyle\scriptscriptstyle}%
{\@@ontopof{#1}{#2}{#3}\scriptscriptstyle\scriptscriptstyle}%
}%
}

\def\@@ontopof#1#2#3#4#5{%
\setbox0=\hbox{$#4#1$}%
\setbox1=\hbox{$#5#2$}%
\setbox2=\hbox{}\ht2=\ht0 \dp2=\dp0 %
\ifdim\wd0>\wd1 %
\setbox1=\hbox to\wd0{\hss\box1\hss}%
\mathord{\rlap{\raise#3\ht0\box1}\box0}%
\else   %
\setbox1=\hbox to.9\wd1{\hss\box1\hss}%
\setbox0=\hbox to\wd1{\hss$#4\relax#1$\hss}%
\mathord{\rlap{\copy0}\raise#3\ht0\box1}%
\fi
}%

\def\lambdabar{\protect\@lambdabar}
\def\@lambdabar{%
\relax
\bgroup
\def\@tempa{\hbox{\raise.73\ht0
\hbox to0pt{\kern.25\wd0\vrule width.5\wd0
height.1pt depth.1pt\hss}\box0}}%
\mathchoice{\setbox0\hbox{$\displaystyle\lambda$}\@tempa}%
{\setbox0\hbox{$\textstyle\lambda$}\@tempa}%
{\setbox0\hbox{$\scriptstyle\lambda$}\@tempa}%
{\setbox0\hbox{$\scriptscriptstyle\lambda$}\@tempa}%
\egroup
}

\def\corresponds{{\lower.2ex\hbox{=}}{\rm\kern-.75em^\triangle}}
\def\succsim{\succ\kern-.9em_\sim\kern.3em}
\def\precsim{\prec\kern-1em_\sim\kern.3em}
\def\slantfrac#1#2{\kern1em^{#1}\kern-.3em/\kern-.1em_{#2}}

\begin{document}                                                                

\baselineskip=15pt

\renewcommand{\arraystretch}{1.5}

\begin{center}
{\Large\bf The Rolling Motion of a Disk on a Horizontal Plane}

\bigskip
Alexander J.~McDonald

\medskip

{\sl Princeton High School, Princeton, New Jersey 08540}

\medskip

Kirk T.~McDonald

\medskip

{\sl Joseph Henry Laboratories, Princeton University,
Princeton, New Jersey 08544}

\medskip

mcdonald@puphep.princeton.edu

\bigskip

(March 28, 2001)

\bigskip

{\large\bf Abstract}
\end{center}

Recent interest in the old problem of the motion of a coin spinning on a 
tabletop has focused on mechanisms
 of dissipation of energy as the angle $\alpha$
of the coin to the table decreases, while the angular velocity $\Omega$ 
of the point of
contact increases.  Following a review of the general equations of motion of
a thin disk rolling without slipping on a horizontal surface, we present 
results of simple experiment on the time dependence of the motion that
indicate the dominant dissipative power loss to be proportional
to the $\Omega^2$ up to and including the last observable cycle.

\section{Introduction}

This classic problem has been treated by many authors, perhaps in greatest 
detail but very succinctly by Routh in article 244 of \cite{Routh}.  
About such problems, Lamb \cite{Lamb} has said, ``It is not that the phenomena, 
though
familiar and often interesting, are held to be specially important, but it 
was regarded rather as a point of honour to shew how the mathematical 
formulation could be effected, even if the solution should prove to be 
impracticable, or difficult of interpretation."  Typically, the role of 
friction was little discussed other than in relation to ``rising''
\cite{Routh,Lamb,HT,Jellett,Gallop,Gray00}.  
The present paper is motivated by
recent discussions \cite{Moffatt,Engh,Moffatt2}
 of friction for small angles of inclination of a
spinning disk to the horizontal supporting surface.

The issues of rolling motion of a disk are introduced in sec.~2 in the
larger context of non-rigid-body motion and rolling motion on
curved surfaces, using the science toy ``Euler's
Disk" as an example.  In our analysis of the motion of a rigid disk on a
horizontal plane we adopt a vectorial approach as advocated by Milne 
\cite{Milne}.  The equations of motion assuming rolling without slipping are
 deduced in sec.~3, steady motion is discussed in secs.~4 and 5, and
oscillation about steady motion is considered in sec.~7.  The case of zero
friction is discussed in secs.~8 and 9, and effects of dynamic
friction are discussed in secs.~6, 10 and 11.  Section 12 presents
a brief summary of the various aspects of the motions discussed in 
secs.~3-11.    

\section{The Tangent Toy ``Euler's Disk"}

An excellent science toy that illustrates the topic of this article is
``Euler's Disk", distributed by Tangent Toy Co.\ \cite{Tangent}.
Besides the disk itself, a base is included that appears to be the key to the
superior performance exhibited by this toy.  The surface of the base is a thin,
curved layer of glass, glued to a plastic backing.  The base rests on 
three support points to minimize rocking.

As the disk rolls on the base, the latter is noticeably deformed.  If the
same disk is rolled on a smooth, hard surface such as a granite surface plate,
the motion dies out more quickly, and rattling sounds are more prominent.
It appears that a small amount of flexibility in the base is important in
damping the perturbations of the rolling motion if long spin times are to be
achieved.

Thus, high-performance rolling motion is not strictly a rigid-body phenomenon.
However, we do not pursue the theme of elasticity further in this paper.

The concave shape of the Tangent Toy base 
helps center the rolling motion of the disk, and speeds up the reduction of 
an initially nonzero radius $b$ to the desirable value of zero.

An analysis of the motion of a disk rolling on a curved surface
 is more complex than that of rolling on a horizontal plane
\cite{sphere}.  For rolling near the bottom of the sphere, the results as
very similar to those for rolling on a plane.  A possibly nonintuitive result
is that a disk can roll stably on the inside of the upper hemisphere of a
fixed sphere, as demonstrated in the motorcycle riding act ``The Globe of
Death'' \cite{globeofdeath}.

\section{The Equations of Motion for Rolling Without Slipping}

\begin{figure}[htp]  % h = here, t = top, b = bottom, p = new page
\begin{center}
\vspace{0.1in}
\includegraphics[width=3.0in]{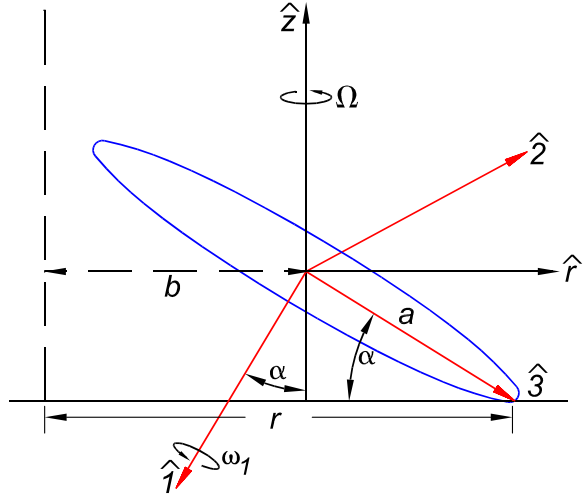}
\parbox{5.5in} % change 5.5in to \hsize for full-width caption
{\caption[ Short caption for table of contents ]
{\label{fig1} A disk of radius $a$ rolls without slipping on a horizontal plane.
The symmetry axis of the disk is called axis $\hat{\bf 1}$, and makes angle
$\alpha$ to the $\hat{\bf z}$ axis, which is vertically upwards, with 
$0 \leq \alpha \leq \pi$.  The line from the center of the
disk to the point of contact with the plane is called axis $\hat{\bf 3}$, 
which makes angle $\alpha$ to the horizontal.  The
horizontal axis $\hat{\bf 2}$ is defined by $\hat{\bf 2} = \hat{\bf 3} \times
\hat{\bf 1}$, and the horizontal axis $\hat{\bf r}$ is defined by $\hat{\bf r}
= \hat{\bf 2} \times 
\hat{\bf z}$.  The angular velocity of the disk about axis $\hat{\bf 1}$
is called $\omega_1$, and the angular velocity of the axes 
$(\hat {\bf 1},\hat {\bf 2},\hat {\bf 3})$ about the vertical is called
$\Omega$.  The motion of the point of contact is instantaneously in a circle of
radius $r$.  The distance from the axis of this motion to the center of mass
of the disk is labeled $b$.
}}
\end{center}
\end{figure}

In addition to the $\hat {\bf z}$ axis which is vertically upwards,
we introduce a right-handed 
coordinate triad of unit vectors $(\hat {\bf 1},\hat {\bf 2},\hat {\bf 3})$ 
related to the geometry of the disk, as shown in Fig.~\ref{fig1}.
Axis $\hat {\bf 1}$ lies along the symmetry axis of the disk. 
Axis $\hat {\bf 3}$ is directed from the center of the disk to the point of
contact with the horizontal plane, and makes angle $\alpha$ to that plane. 
The vector from the center of the disk to the point of contact is then
\begin{equation}
{\bf a} = a\hat {\bf 3}.
\label{e1}
\end{equation}
Axis $\hat {\bf 2} = \hat {\bf 3} \times \hat {\bf 1}$
lies in the plane of the disk, and also in the horizontal plane.
The sense of axis $\hat {\bf 1}$ is chosen  so that the component
$\omega_1$ of the angular velocity vector $\vec\omega$  of the disk about 
this axis is positive.  
 Consequently, axis $\hat{\bf 2}$ points in the direction of the velocity of
the point of contact.  (For the special case where the point of contact does
not move, $\omega_1 = 0$ and analysis is unaffected by the choice of direction
of axis $\hat{\bf 1}$.)

Before discussing the dynamics of the problem, a considerable amount can be
deduced from kinematics.
 The total angular velocity $\vec\omega$ can be thought of as composed of two 
parts,
\begin{equation}
\vec\omega = \vec\omega_{123} + \omega_{\rm rel} \hat{\bf 1},
\label{e2}
\end{equation}
where $\vec\omega_{123}$ is the angular velocity of the triad
$(\hat {\bf 1},\hat {\bf 2},\hat {\bf 3})$, and $\omega_{\rm rel} \hat{\bf 1}$ 
is the angular velocity of the disk relative to that triad; the relative
angular velocity can only have
a component along $\hat{\bf 1}$ by definition.  The angular velocity of the
triad $(\hat {\bf 1},\hat {\bf 2},\hat {\bf 3})$
has component $\dot\alpha$ about the horizontal axis $\hat{\bf 2}$ 
(where the dot indicates differentiation with respect to time), and is defined to have component $\Omega$ about the vertical axis $\hat{\bf z}$.
Since axis $\hat{\bf 2}$ is always horizontal, $\vec\omega_{123}$ has no
component along the (horizonal) axis $\hat{\bf 2} \times \hat{\bf z} 
\equiv \hat{\bf r}$.  Hence, the angular velocity of the triad 
$(\hat {\bf 1},\hat {\bf 2},\hat {\bf 3})$ can be written
\begin{equation}
\vec\omega_{123} = \Omega \hat{\bf z} + \dot\alpha \hat{\bf 2}
= - \Omega \cos \alpha \hat{\bf 1} + \dot\alpha \hat{\bf 2}
  - \Omega \sin\alpha \hat{\bf 3},
\label{e4}
\end{equation}
noting that
\begin{equation}
\hat{\bf z} = - \cos\alpha \hat{\bf 1} - \sin\alpha \hat{\bf 3}, 
\label{e3}
\end{equation}
as can be seen from Fig.~\ref{fig1}.
The time rates of change of the axes are therefore 
\begin{eqnarray}
{d\hat{\bf 1} \over dt} & = & \vec\omega_{123} \times \hat{\bf 1}
= -\Omega \sin\alpha \hat{\bf 2} - \dot\alpha \hat{\bf 3}, 
\label{e5} \\
{d\hat{\bf 2} \over dt} & = & \vec\omega_{123} \times \hat{\bf 2}
= \Omega \sin\alpha \hat{\bf 1} - \Omega \cos\alpha \hat{\bf 3},
= - \Omega \hat{\bf r},
\label{e6} \\
{d\hat{\bf 3} \over dt} & = & \vec\omega_{123} \times \hat{\bf 3}
= \dot\alpha \hat{\bf 1} + \Omega \cos\alpha \hat{\bf 2},
\label{e7}
\end{eqnarray}
where the rotating horizontal axis $\hat{\bf r}$ is related by
\begin{equation}
\hat{\bf r} = \hat{\bf 2} \times \hat{\bf z}
= - \sin\alpha \hat{\bf 1} + \cos\alpha \hat{\bf 3},
\label{e25b}
\end{equation}
whose time rate of change is
\begin{equation}
{d\hat{\bf r} \over dt} = \Omega \hat{\bf 2}.
\label{e250}
\end{equation}

Combining eqs.~(\ref{e2}) and (\ref{e4}) we write the total angular velocity as
\begin{equation}
\vec\omega = \omega_1 \hat{\bf 1} + \dot\alpha \hat{\bf 2}
  - \Omega \sin\alpha \hat{\bf 3},
\label{e8}
\end{equation}
where
\begin{equation}
\omega_1 = \omega_{\rm rel} - \Omega \cos \alpha. 
\label{e9}
\end{equation}

The (nonholonomic)
constraint that the disk rolls without slipping relates the velocity of
the center of mass to the angular velocity vector $\vec\omega$ of the disk.
In particular, the instantaneous velocity of the point contact of the disk with
the horizontal plane is zero,
\begin{equation}
{\bf v}_{\rm contact} = {\bf v}_{\rm cm} + \vec\omega \times {\bf a} = 0.
\label{e10}
\end{equation}
Hence,
\begin{equation}
{\bf v}_{\rm cm} = {d {\bf r}_{\rm cm} \over dt} 
= a\hat {\bf 3} \times \vec\omega
= - a \dot\alpha \hat{\bf 1} + a \omega_1 \hat{\bf 2}, 
\label{e11}
\end{equation}
using eqs.~(\ref{e1}) and (\ref{e8}).  

Additional kinematic relations
can be deduced by noting that the point of contact
between the disk and the horizontal plane can always be considered as moving
instantaneously in a circle whose radius vector we define as ${\bf r} = r
\hat{\bf r}$ with $r \ge 0$, as shown in Fig.~\ref{fig1}, and whose center
is defined to have position $x_A \hat{\bf x} + y_A \hat{\bf y}$
 where $\hat{\bf x}$
and $\hat{\bf y}$ are fixed horizontal unit vectors in the lab frame.  
Then, the center of mass of the disk has position
\begin{equation}
{\bf r}_{\rm cm} = x_A \hat{\bf x} + y_A \hat{\bf y} +
r \hat{\bf r} - a \hat{\bf 3},
\label{e10a}
\end{equation}
and
\begin{equation}
{d {\bf r}_{\rm cm} \over dt} = \dot x_A \hat{\bf x} + \dot y_A \hat{\bf y}
+ \dot r \hat{\bf r} 
 - a \dot\alpha \hat{\bf 1} + (r - a \cos\alpha) \Omega \hat{\bf 2}.
\label{e10b}
\end{equation}
In the special case of steady motion, $\dot x_A = \dot y_A = \dot r =
\dot \alpha = 0$, eqs.~(\ref{e11}) and (\ref{e10b}) combine to give
\begin{equation}
\omega_{1} = {b \over a} \Omega,
\label{e11a}
\end{equation}
where
\begin{equation}
b = r - a \cos\alpha
\label{e11b}
\end{equation}
is the horizontal
distance from the axis of the circular motion to the center of 
mass of the disk.  Thus, for steady motion
%the center of mass (and the point of contact) moves in a circle, and
%we learn that the condition of rolling without
%slipping requires the point of contact to move in a circle of constant
%radius, even though the angle $\alpha$ (and angular velocity $\Omega$) may
%vary with time.  Also, 
the ``spin'' angular velocity $\omega_1$ is related
to the ``precession" angular velocity $\Omega$ according to 
eq.~(\ref{e11a}).  
While $\omega_1$ is defined to be nonnegative, length $b$ can be negative if
$\Omega$ is negative as well.
%We could use either $\omega_1$ or $b$ %(or $r$)
% as one of the basic variables of the
%problem.  For now, we continue to use $\omega_1$, as we wish to include the
%special cases of $b = 0$ and $\infty$ in the general analysis.

Except for axis $\hat {\bf 1}$, the rotating axes 
$(\hat {\bf 1},\hat {\bf 2},\hat {\bf 3})$ are not body axes,
but the inertia tensor $I_{ij}$ is diagonal with respect to them in view of 
the symmetry of the disk.  We write
\begin{equation}
I_{11} = 2kma^2, \qquad I_{22} = kma^2 = I_{33},
\label{e12}
\end{equation}
which holds for any thin, circularly symmetric disk according to the
perpendicular axis theorem; $k = 1/2$ for a disk with
mass $m$ concentrated at the rim, $k = 1/4$ for a uniform disk, \etc\ \
The angular momentum ${\bf L}_{\rm cm}$ of the disk with respect to its center
of mass can now be written as
\begin{equation}
{\bf L}_{\rm cm} = \vec{\vec I} \cdot \vec \omega = 
kma^2 (2 \omega_1 \hat {\bf 1} 
+ \dot\alpha \hat{\bf 2} 
- \Omega \sin\alpha \hat{\bf 3}).
\label{e13}
\end{equation}

Turning at last to the dynamics of the rolling disk, we suppose that the only
forces on it are $- m g \hat{\bf z}$ due to gravity and  
{\bf F} at the point of contact with the horizontal plane.  For now, we ignore
rolling friction and friction due to the air surrounding the disk.  

The equation of motion for the position ${\bf r}_{\rm cm}$ of
 the center of mass of the disk is then
\begin{equation}
m {d^2 {\bf r}_{\rm cm} \over dt^2} = {\bf F} - mg \hat {\bf z}.
\label{e14}
\end{equation}
The torque equation of motion for the angular momentum ${\bf L}_{\rm cm}$ 
about the center of mass is
\begin{equation}
{d {\bf L}_{\rm cm} \over dt} = {\bf N}_{\rm cm} = {\bf a} \times {\bf F}.
\label{e15}
\end{equation}
We eliminate the unknown force ${\bf F}$ in eq.~(\ref{e15}) via 
eqs.~(\ref{e1}) and (\ref{e14}) to find
\begin{equation}
{1 \over ma} {d {\bf L}_{\rm cm} \over dt} 
+ {d^2 {\bf r}_{\rm cm} \over dt^2} \times \hat {\bf 3}
= g\hat {\bf 3} \times \hat {\bf z}.
\label{e17}
\end{equation}
This can be expanded using eqs.~(\ref{e3}), (\ref{e5})-(\ref{e7}), (\ref{e11})
and (\ref{e13}) to yield the $\hat{\bf 1}$, $\hat{\bf 2}$ and $\hat{\bf 3}$
components of the equation of motion,
\begin{eqnarray}
(2k + 1) \dot\omega_1 + \dot\alpha \Omega \sin\alpha & = & 0,
\label{e18} \\
k \Omega^2 \sin\alpha \cos\alpha
 + (2k + 1) \omega_1 \Omega \sin\alpha - (k + 1) \ddot\alpha
 & = &  {g \over a} \cos\alpha,
\label{e19} \\
\dot\Omega \sin\alpha + 2 \dot\alpha \Omega \cos\alpha + 2 \omega_1 \dot\alpha
& = & 0.
\label{e20}
\end{eqnarray}
These equations agree with those of sec.~244 of Routh \cite{Routh}, noting
that his $A$, $C$, $\theta$, $\dot\psi$ and $\omega_3$ are expressed as
$k a^2$, $2 k a^2$, $\alpha$, $\Omega$ and $- \omega_1$, respectively, in our
notation.

Besides the coordinates ($x_A,y_A)$ of the center of motion,
we can readily identify only one other constant of the motion, 
% \cite{Olsson}, 
the total energy
\begin{eqnarray}
E & = & T + V =  {1 \over 2} m v_{\rm cm}^2 + {1 \over 2} \vec \omega \cdot 
\vec{\vec I} \cdot \vec \omega + m g z
\nonumber \\
& = & {m a^2 \over 2} \left[ (2 k + 1) \omega_1^2 + (k + 1) \dot\alpha^2 +
k \Omega^2 \sin^2\alpha + {2 g \over a} \sin\alpha \right].
\label{e20a}
\end{eqnarray}
The time derivative of the energy is consistent with the
equations of motion (\ref{e18})-(\ref{e20}), but does not provide any 
independent information.

\section{Steady Motion}

For steady motion, $\dot\alpha = \ddot\alpha = \dot\Omega = \dot\omega_1 = 0$,
and we define $\alpha_{\rm steady} = \alpha_0$, 
$\Omega_{\rm steady} = \Omega_0$ and $\omega_{1,\rm steady}
= \omega_{10}$.  The equations of motion
(\ref{e18}) and (\ref{e20}) are now trivially satisfied, and eq.~(\ref{e19})
becomes
\begin{equation} 
k \Omega_0^2 \sin\alpha_0 \cos\alpha_0 + (2k + 1) \omega_{10} \Omega_0
 \sin\alpha_0
= {g \over a} \cos\alpha_0,
\label{e21}
\end{equation}

A special case of steady motion is $\alpha_0 = \pi / 2$, corresponding to the
plane of the disk being vertical.  In this case, eq.~(\ref{e21}) requires that
$\omega_{10} \Omega_0 = 0$.  
If $\Omega_0 = 0$, the disk rolls along a straight line and $\omega_{10}$ is 
the rolling angular velocity.
If $\omega_{10} = 0$, the disk spins in place about the vertical axis
with angular velocity $\Omega_0$.

For $\alpha_0 \neq \pi/2$, the angular velocity $\Omega_0 \hat{\bf z}$ of the 
axes about the vertical must be nonzero.  
We can then replace $\omega_{10}$ by the radius $b$ of the horizontal circular
motion of the center of mass using eqs.~(\ref{e11a})-(\ref{e11b}):
\begin{equation}
\omega_{10} = {b \over a} \Omega_0 = \Omega_0 \left( {r \over a} - \cos\alpha_0
\right).
\label{e22}
\end{equation}
Inserting this in (\ref{e21}), we find
\begin{equation} 
\Omega_0^2 = {g \cot\alpha_0 \over k a \cos\alpha_0 + (2k + 1) b}
=  {g \cot\alpha_0 \over (2k + 1) r - (k + 1) a \cos\alpha_0 }.
\label{e23}
\end{equation}

For $\pi / 2 < \alpha_0 < \pi$ the denominator of eq.~(\ref{e23}) is positive,
since $r$ is positive by definition, but the numerator is negative.  Hence,
$\Omega_0$ is imaginary, and steady motion is not possible in this quadrant
of angle $\alpha_0$.

For $0 < \alpha_0 < \pi / 2$, $\Omega_0$ is real and steady motion is possible
so long as
\begin{equation}
b > - {a k \cos\alpha_0 \over 2 k + 1}.
\label{e23a}
\end{equation}
In addition to the commonly observed case of $b > 0$, steady motion 
is possible with small negative values of $b$

A famous special case is when $b = 0$, and the center of mass of the disk is at
rest.  Here, eq.~(\ref{e23}) becomes
\begin{equation} 
\Omega_0^2 = {g \over a k \sin\alpha_0},
\label{e24}
\end{equation}
and $\omega_{10} = 0$ according to eq.~(\ref{e22}), so that
\begin{equation}
\omega_{\rm rel} = \Omega_0 \cos\alpha_0,
\label{e25}
\end{equation}
recalling eq.~(\ref{e9}).  Also, the total angular velocity becomes simply
$\vec\omega = - \Omega_0 \sin\alpha_0 \hat{\bf 3}$ according to eq.~(\ref{e8}),
so the instantaneous axis of rotation is axis $ {\bf 3}$ which contains the
center of mass and the point of contact, both of which are instantaneously at
rest.   

\section{Shorter Analysis of Steady Motion with $b = 0$}

The analysis of a spinning disk (for example, a coin) whose center is at
rest can be shortened
considerably by noting at the outset that in this case
axis $\hat{\bf 3}$ is the instantaneous axis of rotation.
Then, the angular velocity is $\vec\omega = \omega \hat{\bf 3}$, and
the angular momentum is simply
\begin{equation}
{\bf L} = I_{33} \omega \hat{\bf 3} = k m a^2 \omega \hat{\bf 3}.
\label{e26}
\end{equation}
Since the center of mass is at rest, the contact force {\bf F} is just
$mg \hat{\bf z}$, so the torque about the center of mass is 
\begin{equation}
{\bf N} = a \hat{\bf 3} \times mg \hat{\bf z} = {d{\bf L} \over dt}.
\label{e27}
\end{equation}
We see that the equation of motion for {\bf L} has the form
\begin{equation}
{d{\bf L} \over dt} = \vec\Omega_0 \times {\bf L},
\label{e28}
\end{equation}
where
\begin{equation}
\vec\Omega_0 = - {g \over a k \omega} \hat{\bf z}.
\label{e29}
\end{equation}
Thus, the angular momentum, and the coin precesses about the vertical at rate
$\Omega_0$.

A second relation between $\vec\omega$ and $\vec\Omega_0$ is obtained from
eqs.~(\ref{e2}) and (\ref{e4})
by noting that $\vec\omega_{123} = \vec\Omega_0$, so that
\begin{equation}
\vec\omega = (- \Omega_0 \cos\alpha_0 + \omega_{\rm rel} )\hat{\bf 1} 
- \Omega_0 \sin\alpha_0 \hat{\bf 3} = \omega \hat{\bf 3},
\label{e30}
\end{equation}
using eq.~(\ref{e3}).  Hence,
\begin{equation}
\omega = - \Omega_0 \sin\alpha_0,
\label{e31}
\end{equation}
and the angular velocity $\omega_1$ about the symmetry axis vanishes, so that
\begin{equation}
\omega_{\rm rel} = \Omega_0 \cos\alpha_0.
\label{e32}
\end{equation}
Combining eqs.~(\ref{e29}) and (\ref{e31}), we again find that
\begin{equation} 
\Omega_0^2 = {g \over a k \sin\alpha_0},
\label{e33}
\end{equation}

As $\alpha_0$ approaches zero, the angular velocity
of the point of contact becomes very large, and one hears a high-frequency
sound associated with the spinning coin.  However, a prominent aspect of what
one sees is the rotation of the figure on the face of the coin, whose angular
velocity $\Omega_0 - \omega_{\rm rel} = \Omega_0 (1 - \cos\alpha_0)$ approaches
zero.   The total angular velocity $\omega$ also vanishes as $\alpha_0 \to 0$.

\section{Radial Slippage During ``Steady'' Motion}

The contact force {\bf F} during steady motion at a small angle $\alpha_0$
is obtained from eqs.~(\ref{e6}), (\ref{e11}), (\ref{e14}), (\ref{e22}) and
(\ref{e24}) as
\begin{equation}
{\bf F} = mg \hat{\bf z} - {b \over a k \sin\alpha_0} mg \hat{\bf r}.
\label{e25a}
\end{equation}
%where axis $\hat{\bf r}$ is defined by eq.~(\ref{e25b}).
The horizontal component of force {\bf F} is due to static friction at
the point of contact.  The coefficient $\mu$ of friction must therefore
satisfy
\begin{equation}
\mu \ge {\abs{b} \over a k \sin\alpha_0},
\label{e25c}
\end{equation}
otherwise the disk will slip in the direction opposite to the radius vector
{\bf b}.  Since
coefficient $\mu$ is typically one or less, slippage will occur whenever
$a k \sin\alpha_0 \lsim \abs{b}$.  As the disk loses energy and angle $\alpha$ 
decreases, the slippage will reduce $\abs{b}$ as well.  
The trajectory of the center of the disk
will be a kind of inward spiral leading toward $b = 0$ for small $\alpha$.

If distance $b$ is negative, it must obey $\abs{b} < a k \cos\alpha_0 /
(2 k + 1)$ according to eq.~(\ref{e23a}).  In this case, eq.~(\ref{e25c})
becomes
\begin{equation}
\mu \ge {\cot\alpha_0 \over 2 k + 1},
\label{e25d}
\end{equation}
which could be satisfied for a uniform disk only for 
$\alpha_0 \gsim \pi / 3$.  Motion with negative $b$ is likely to be observed
only briefly before large radial slippage when $\alpha_0$ is large reduces
$b$ to zero.

Once $b$ is zero, the contact force is purely vertical, according to 
eq.~(\ref{e25a}).  Surprisingly, the condition of rolling without slipping
can be maintained in this special case without any friction at the point of contact.  Hence, an analysis of the motion with $b = 0$ could be made
with the assumption of zero friction, as discussed in secs.~8-9.  
In practice,
there will always be some friction, aspects of which are further discussed in 
secs.~10-11.  From the argument here, we see that if $b = 0$, there is
no frictional force to oppose the radial slippage that accompanies a change in angle $\alpha$.

\section{Small Oscillations about Steady Motion}

We now consider oscillations at angular frequency 
$\varpi$ about steady motion, assuming that the disk rolls without 
slipping.  We suppose that $\alpha$, $\Omega$ and $\omega_{1}$ have the form
\begin{eqnarray}
\alpha & = & \alpha_0 + \epsilon \cos\varpi t,
\label{e41} \\
\Omega & = & \Omega_0 + \delta \cos\varpi t,
\label{e42} \\
\omega_1 & = & \omega_{10} + \gamma \cos\varpi t,
\label{e43}
\end{eqnarray}
where $\epsilon$, $\delta$ and $\gamma$ are small constants.  Inserting these
in the equation of motion (\ref{e20}) and equating terms of first order of
smallness, we find that
\begin{equation}
\delta = - {2 \epsilon \over \sin\alpha_0} (\Omega_0 \cos\alpha_0 + \omega_{10}).
\label{e44}
\end{equation}
From this as well as from eq.~(\ref{e41}), we see that $\epsilon / 
\sin\alpha_0 \ll 1$ for small oscillations.
Similarly, eq.~(\ref{e18}) leads to
\begin{equation}
\gamma = - \epsilon {\Omega_0 \sin\alpha_0 \over 2 k + 1},
\label{e45}
\end{equation}
and eq.~(\ref{e19}) leads to
\begin{eqnarray}
\epsilon \varpi^2 (k + 1) & = & 
- (2 k + 1) (\epsilon \omega_{10} \Omega_0 \cos\alpha_0
+ \gamma \Omega_0 \sin\alpha_0  + \delta \omega_{10} \sin\alpha_0)
+ \epsilon k \Omega_0^2 (1 - 2 \cos^2\alpha_0)
\nonumber \\
& & \ - 2 \delta k \Omega_0 \sin\alpha_0 \cos\alpha_0
- \epsilon  {g \over a} \sin\alpha_0.
\label{e46}
\end{eqnarray}

Combining eqs.~(\ref{e44})-(\ref{e46}), we obtain
\begin{eqnarray}
\varpi^2 (k + 1) & = & 
\Omega_0^2 (k (1 + 2 \cos^2\alpha_0) + \sin^2\alpha_0)
+ (6k + 1) \omega_{10} \Omega_0 \cos\alpha_0
\nonumber \\
& & + 2 (2 k + 1) \omega^2_{10}
- {g \over a} \sin\alpha_0, 
\label{e47}
\end{eqnarray}
which agrees with Routh \cite{Routh}, noting that our $k$, $\Omega_0$, and
$\omega_{10}$ are his $k^2$, $\mu$, and $- n$.

For the special case of a wheel rolling in a straight line, $\alpha_0 = \pi/2$,
$\Omega_0 = 0$, and 
\begin{equation}
\varpi^2 (k + 1) =  2 (2 k + 1) \omega^2_{10} - {g \over a}. 
\label{e48}
\end{equation}
The rolling is stable only if
\begin{equation}
  \omega^2_{10} >  {g \over  2 (2 k + 1) a}. 
\label{e49}
\end{equation}

Another special case is that of a disk spinning about a vertical diameter,
for which $\alpha_0 = \pi / 2$ and $\omega_{10}$ and $b$ are zero.  Then,
eq.~(\ref{e47}) indicates that the spinning is stable only for
\begin{equation}
\abs{\Omega_0} > \sqrt{g \over a (k + 1)},
\label{e49a}
\end{equation}
which has been called the condition
for ``sleeping''.  Otherwise, angle $\alpha$ decreases when perturbed,
and the motion of the disc becomes that of the more general case.
Further discussion of this special case is given in the following section.

Returning to the general analysis of eq.~(\ref{e47}), we eliminate $\omega_{10}$
using eq.~(\ref{e22}) and replace the term $(g / a) \sin\alpha_0$ via
eq.~(\ref{e23}) to find
\begin{eqnarray}
{\varpi^2 \over \Omega_0^2} (k + 1) & = & 
 3 k \cos^2\alpha_0 + \sin^2\alpha_0
+ {b \over a} \left( (6k + 1) \cos\alpha_0 
                  - (2 k + 1) {\sin^2\alpha_0 \over \cos\alpha_0} \right)
\nonumber \\
& & \qquad  +\  2 {b^2 \over a^2} (2 k + 1). 
\label{e50}
\end{eqnarray}

The term in eq.~(\ref{e50}) in large parentheses is negative for $\alpha_0
> \tan^{-1} \sqrt{(6 k + 1) / (2 k + 1)}$, which is about $60^\circ$ for a
uniform disk.  Hence,  for positive $b$ the motion is unstable for large
$\alpha_0$, and the disk will appear fall over quickly into a rolling
motion with $\alpha_0 \lsim 60^\circ$, after which $\alpha_0$ will
decrease more slowly due to the radial slippage discussed in sec.~5, until
$b$ becomes very small.  The subsequent motion at small $\alpha_0$ is
considered further in sec.~11.

The motion with negative $b$ is always stable against small oscillations,
but the radial slippage is large as noted in sec.~5.

For motion with $b \ll a$, such as for a spinning coin whose center is nearly
fixed, the frequency of small oscillation is given by
\begin{equation}
{\varpi \over \Omega_0} = \sqrt{3 k \cos^2\alpha_0 + \sin^2\alpha_0 
\over k + 1}. 
\label{e51}
\end{equation}
For small angles this becomes
\begin{equation}
{\varpi \over \Omega_0} \approx \sqrt{3 k \over k + 1}. 
\label{e52}
\end{equation}
For a uniform disk with $k = 1/4$, the frequency $\varpi$ of small oscillation 
approaches $\sqrt{3/5} \Omega_0 = 0.77 \Omega_0$, while for a hoop with 
$k = 1/2$, $\varpi \to \Omega_0$ as $\alpha_0 \to 0$.

The effect of this small oscillation of a spinning coin is to produce a kind of 
rattling sound during which the frequency sounds a bit ``wrong".  This may be
particularly noticeable if a surface imperfection suddenly excites the 
oscillation to a somewhat larger amplitude.

The radial slippage of the point of contact discussed in sec.~5 
will be enhanced
by the rattling, which requires a larger peak frictional force to
maintain slip-free motion.
 
As angle $\alpha_0$ approaches zero, the slippage keeps
the radius $b$ of order $ a \sin\alpha_0$.  For small $\alpha_0$, 
$b \approx \alpha_0 a$ and
eq.~(\ref{e50}) gives the frequency of small oscillation as
\begin{equation}
\varpi \approx \Omega_0 \sqrt{3 k +  (6 k + 1) \alpha_0 \over k + 1}.
\label{e53}
\end{equation}
For a uniform disk, $k = 1/4$, and eq.~(\ref{e53}) gives
\begin{equation}
\varpi \approx \Omega_0 \sqrt{3 + 10 \alpha_0 \over 5}.
\label{e54}
\end{equation}
When $\alpha_0 \approx 0.2$ rad, the oscillation and rotation frequencies are
nearly identical, at which time 
a very low frequency beat can be discerned in the
nutations of the disk.  Once $\alpha_0$ drops below about 0.1 rad, the
low-frequency nutation disappears and the disk settles into a motion in which
the center of mass hardly appears to move, and the rotation frequency
$\Omega_0 \approx \sqrt{g / a k \alpha_0}$ grows very large.

For a hoop ($k = 1/2$), the low-frequency beat will be prominent for angles
$\alpha$ near zero.

\section{Disk Spinning About a Vertical Diameter}

When a disc is spinning about a vertical diameter the condition of contact
with the horizontal surface is not obviously that of rolling without
slipping, which requires nonzero static friction.
Olsson has suggested that there is zero friction between the disk and the
surface in this case \cite{Olsson}.

If there is no friction, all forces on the disc are vertical.  Then, the
center of mass moves only vertically, and there is no vertical torque 
component about the center of mass, so the vertical component $L_z$ of
angular momentum is constant.

The equations of motion in the absence of friction can be found by the
method of sec.~3, writing the position of the center of mass as
\begin{equation}
{\bf r}_{\rm cm} = a \sin\alpha \hat{\bf z}.
% \qquad  {\bf v}_{\rm cm} = a \dot\alpha \cos\alpha \hat{\bf z}.
\label{e401}
\end{equation}
Using this in eq.~(\ref{e17}), 
 the $\hat{\bf 1}$, $\hat{\bf 2}$ and $\hat{\bf 3}$
components of the equation of motion are
\begin{eqnarray}
\dot\omega_1 & = & 0,
\label{e402} \\
(k \Omega^2 + \dot\alpha^2) \sin\alpha \cos\alpha
 + 2k \omega_1 \Omega \sin\alpha - (k + \cos^2\alpha) \ddot\alpha
 & = &  {g \over a} \cos\alpha,
\label{e403} \\
\dot\Omega \sin\alpha + 2 \dot\alpha \Omega \cos\alpha + 2 \omega_1 \dot\alpha
& = & 0.
\label{e404}
\end{eqnarray}
According to eq.~(\ref{e402}),  the angular velocity $\omega_1$ about the
symmetry axis of the disk is constant.  Then, eq.~(\ref{e404}) can be
multiplied by $k m a^2 \sin\alpha$ and integrated to give
\begin{equation}
k m a^2 (\Omega \sin^2\alpha - 2 \omega_1 \cos\alpha)
 = L_z =\ {\rm constant},
\label{e405}
\end{equation}
recalling eq.~(\ref{e13}).

In the case of motion of a disk with no friction we find five constants of
the motion, $x_{\rm cm}$, $y_{\rm cm}$, $\omega_1$, $L_z$ and the total
energy $E$, in contrast to the
case of rolling without slipping in which the only (known) constants of the
motion are the energy $E$ and the coordinates ($x_A,y_A$) of the
center of motion.

For spinning about a vertical diameter, $\alpha = \pi /2$ and $\omega_1 = 0$.
For small perturbations about this motion we write $\alpha = \pi / 2 - 
\epsilon$, and for small $\epsilon$ eq.~(\ref{e403}) becomes
\begin{equation}
\ddot\epsilon + \left( \Omega^2 - {g \over a k} \right) \epsilon = 0.
\label{e406}
\end{equation}
Hence, in the case of no friction, spinning about a vertical diameter is
stable for
\begin{equation}
\abs{\Omega} > \sqrt{g \over a k}\, .
\label{e407}
\end{equation}
For a uniform disk with $k = 1/4$, this stability condition is that
$\abs{\Omega} > 2\sqrt{g/a}$.

In contrast, the stability condition (\ref{e49a}) for a uniform disk that
rolls without slipping is that $\abs{\Omega} > 2\sqrt{g/5a} \approx
0.9 \sqrt{g/a}$.

As the stability conditions (\ref{e49a}) and (\ref{e407}) differ by more
than a factor of two for a uniform disk, there is hope of distinguishing
between them experimentally.  

We conducted several tests in which a U.S.\
quarter dollar was spun initially about a vertical diameter on a vinyl floor, 
on a sheet of glossy paper, and on the glass surface of the base of the
Tangent Toy Euler's Disk.  (The Euler's Disk is so thick that when spun about
a vertical diameter it comes to rest without falling over.)

We found it essentially impossible to spin a coin such that there is
no motion of its center of mass.  Rather, the center of mass moves slowly
in a spiral before the coin falls over into the ``steady'' motion with
small $b$ described in sec.~5.  A centripetal force is required for such
spiral motion, and so friction cannot be entirely neglected.  The
occasional observation of ``rising'', as discussed further in sec.~10,
is additional evidence for the role of friction in nearly vertical
spinning.

Analysis of frames taken with a digital video camera \cite{Panasonic} at
30 frames per second with exposure time 1/8000 s did
not reveal a sharp transition from spinning of a coin nearly vertically
about a diameter to the settling motion of sec.~5, but in our judgment 
the transition point for $\Omega$ in several data sets was in the range 
1.5-3$\sqrt{g/a}$.  This suggests that during the spinning about a nearly
 vertical 
diameter friction plays only a small role, as advocated by Olsson 
\cite{Olsson}.

\section{Small Oscillations About Steady Motion with No Friction}

It is interesting to pursue the consequences of the equations of motion
(\ref{e402})-(\ref{e404}), deduced assuming no friction,
when angle $\alpha$ is different from $\pi / 2$.
For motion that has evolved from $\alpha = \pi / 2$ initially, we expect the
constant $\omega_1$ to be zero still.  Then, eq.~(\ref{e403}) indicates that
the value of $\Omega_0$ for steady motion at angle $\alpha_0$ is the same as
that of eq.~(\ref{e24}).  This was anticipated in sec.~5, where it was noted
that for $b = 0$ no friction is required to enforce the condition of 
rolling without slipping.

We consider small oscillations about steady motion at angle $\alpha_0$ of
the form
\begin{eqnarray}
\alpha & = & \alpha_0 + \epsilon \cos\varpi t,
\label{e408} \\
\Omega & = & \Omega_0 + \delta \cos\varpi t,
\label{e409}
\end{eqnarray}
where $\epsilon$ and $\delta$ are small constants.  Inserting these
in the equation of motion (\ref{e404}) and equating terms of first order of
smallness, we find that
\begin{equation}
\delta = - 2 \epsilon \Omega_0 \cot\alpha_0,
\label{e410}
\end{equation}
which is the same as eq.~(\ref{e44}) with $\omega_1 = 0$, since 
eqs.~(\ref{e20}) and (\ref{e404}) are the same.
%From this as well as from eq.~(\ref{e41}), we see that $\epsilon / 
%\sin\alpha_0 \ll 1$ for small oscillations.
Similarly, eq.~(\ref{e403}) leads to
\begin{eqnarray}
\epsilon \varpi^2 (k + \cos^2\alpha_0) & = & 
\epsilon k \Omega_0^2 (1 - 2 \cos^2\alpha_0)
- 2 \delta k \Omega_0 \sin\alpha_0 \cos\alpha_0
- \epsilon  {g \over a} \sin\alpha_0.
\label{e411}
\end{eqnarray}
Combining eqs.~(\ref{e410})-(\ref{e411}), we obtain
\begin{equation}
\varpi^2 (k + \cos^2\alpha_0) =  
k \Omega_0^2 (1 + 2 \cos^2\alpha_0) - {g \over a} \sin\alpha_0
= 3 k \Omega_0^2 \cos^2\alpha_0,
\label{e412}
\end{equation}
using eq.~(\ref{e24}).  The ratio of the frequency $\varpi$ of small 
oscillations to the frequency $\Omega_0$ of rotation about the vertical
axis for $\alpha_0 < \pi / 2$ is
\begin{equation}
{\varpi \over \Omega_0} = \sqrt{3 k \over k + \cos^2\alpha_0} \cos\alpha_0,
\label{e413}
\end{equation}
which differs somewhat from the result (\ref{e51}) obtained assuming
rolling without slipping.  For very small $\alpha_0$, both eq.~(\ref{e51}) and
(\ref{e413}) take on the same limiting value (\ref{e52}).

Because of the similarity of the results for small oscillations about
steady motion with $b = 0$ for either assumption of no friction or rolling without slipping, it will be hard to distinguish experimentally which condition is the more realistic, but the distinction is of little consequence.

\section{``Rising'' of a Rotating Disk When Nearly Vertical ($\alpha \approx
\pi / 2$)}

A rotating disk can exhibit ``rising" when launched with
spin about a nearly vertical diameter, provided there is slippage at the
point of contact with the horizontal plane.  That is, the plane of the disc may
rise first towards the vertical, before eventually
falling towards the horizontal. 

The rising of tops appears to have been considered by Euler, but rather
inconclusively.  The present explanation based on sliding friction  can be 
traced to a note by ``H.T.'' in 1839 \cite{HT}.  

Briefly, we consider motion
that is primarily rotation about a nearly vertical diameter.
The  angular velocity about the vertical is $\Omega > \sqrt{g/a k}$,
large enough so that ``sleeping'' at the vertical is possible in the
absence of friction.  
The needed sliding friction depends on angular velocity component $\omega_1
= b \Omega / a$ being nonzero, which
implies that the center of mass moves in a circle of radius $b \ll a$
in the present case.  Then, $\omega_1 \ll \Omega$, and the angular momentum
(\ref{e13}) is  ${\bf L} \approx - \Omega \hat{\bf 3}$, which is
almost vertically upwards (see Fig.~\ref{fig1}).  
Rising depends on slippage of
the disk at the point of contact such that the lowermost point on the disk
is not at rest but moves with velocity $- \epsilon a \omega_1 \hat{\bf 2}$,
which is opposite to the direction of motion of the center of mass.  
Corresponding to this slippage, the horizontal surface exerts friction 
$F_s \hat{\bf 2}$ on the disk, with $F_s > 0$.  The related torque,
${\bf N}_s = a \hat{\bf 3} \times F_s \hat{\bf 2} = - a F_s \hat{\bf 1}$,
pushes the angular momentum towards the vertical, and the center of mass of
the disk rises.

The torque needed for rising exists in principle even for a disk of zero
thickness, provided there is sliding friction at the point of contact.

The most dramatic form of rising motion is that of a ``tippe'' top, which
has recently been reviewed by Gray and Nickel \cite{Gray00}.

\section{Friction at Very Small $\alpha$}

In practice, the motion of a spinning disk appears to cease rather abruptly for
a small value of the angle $\alpha$, corresponding to large
precession angular velocity $\Omega$.  If the motion continued, the velocity
$\Omega a$ of the point of contact would eventually exceed the speed of sound.

This suggests that air friction may play a role in the motion at very small
$\alpha$, as has been discussed recently by Moffatt 
\cite{Moffatt,Engh,Moffatt2}.

When the rolling motion ceases, the disk seems to float for a moment, and then
settle onto the horizontal surface.  It appears that the upward contact force
${\bf F}_z$ vanished, and the disk lost contact with the surface.  From 
eqs.~(\ref{e11}) and (\ref{e14}), we see that for small $\alpha$,
${\bf F}_z \approx mg + m a \ddot \alpha$.  Since the height of the center of
mass
above the surface is $h \approx a \alpha$ for small $\alpha$, we recognize that
the disk loses contact with the surface when the center of mass is falling with
acceleration $g$.

Moffatt invites us to relate the power $P$ dissipated by friction to the rate of
change $dU/dt$ of total energy of the disk.  For a disk moving with $b = 0$ at
a small angle $\alpha(t)$,
\begin{equation}
U = {1 \over 2} m \dot h^2 + {1 \over 2} I_{33} \omega^2 + m g h
\approx {1 \over 2} m a^2 \dot\alpha^2 + {3 \over 2} m a g \alpha,
\label{e101}
\end{equation}
using eq.~(\ref{e31}) and assuming that eq.~(\ref{e33}) holds adiabatically.
Then,
\begin{equation}
P = {d U \over dt} \approx  m a^2 \dot\alpha \ddot\alpha 
+ {3 \over 2} m a g \dot\alpha
\approx {5 \over 2} m a g \dot\alpha,
\label{e102}
\end{equation}
where the second approximation holds when ${\bf F}_z \approx 0$ and
$m a \ddot\alpha \approx m g$.

For the dissipation of energy we need a model.  First, we consider rolling 
friction, taken to be the effect of inelastic collisions between the disk and
the horizontal surface.  For example, suppose the surface has small bumps 
with average spacing $\delta$ and average height $h =\epsilon \delta$.  
We also suppose that the disk dissipates energy 
$m g h = m g \epsilon \delta$ when passing over a 
bump.  The time taken for the rotating disk to pass over a bump is 
$\delta / a \Omega$ (at small $\alpha$), so the
rate of dissipation of energy to rolling friction is 
\begin{equation}
P = - {m g \epsilon \delta \over \delta / a \Omega} 
= - \epsilon m a g \Omega.
\label{e103}
\end{equation}
A generalized form of velocity-dependent friction could be written as
\begin{equation}
P = - \epsilon m a g \Omega^\beta,
\label{e103a}
\end{equation}
where the drag force varies with (angular) velocity as $\Omega^{\beta - 1}$.
A rolling frictional force proportional to the velocity of the contact point
corresponds to $\beta = 2$; an air drag force proportional to the square of the
velocity corresponds to $\beta = 3$.
The model of Moffatt \cite{Moffatt} emphasizes the viscous shear of
the air between the disk and the supporting horizontal surface, and 
corresponds to $\beta = 4$.  A revised version of Moffatt's model reportedly 
\cite{Moffatt2} corresponds to $\beta = 2.5$.

Equating the frictional power loss to the rate of change (\ref{e102}) of the 
energy of the disk, we find
\begin{equation}
\dot\alpha = - {2 \epsilon \over 5} \Omega^\beta 
\approx - {2 \epsilon \over 5} \left( {g \over ak} \right)^{\beta/2} 
{1 \over \alpha^{\beta/2}}\, ,
\label{e104}
\end{equation}
using $\Omega^2 \approx g / a k \alpha$ from eq.~(\ref{e24}), 
which integrates to give
\begin{equation}
\alpha^{(\beta + 2)/2} = {\epsilon (\beta + 2) \over 5} 
\left( {g \over ak} \right)^{\beta/2} (t_0 - t),
\label{e105}
\end{equation}
and
\begin{equation}
\alpha = \left( {\epsilon (\beta + 2) \over 5} \right)^{2/(\beta + 2)}
 \left( {g \over ak} \right)^{\beta/(\beta + 2)} (t_0 - t)^{2/(\beta + 2)}.
\label{e106}
\end{equation}
In this model, the angular velocity $\Omega$ obeys
\begin{equation}
\Omega = \left( {5 g / \epsilon (\beta + 2) a k \over t_0 - t} 
\right)^{1/(\beta + 2)}
\equiv \left( {C \over t_0 - t} 
\right)^{1/(\beta + 2)},
\label{e107}
\end{equation}
which exhibits what is called by Moffatt a ``finite-time singularity" at time
$t_0$ \cite{Moffatt} for any value of $\beta$ greater than $-2$.

A premise of this analysis is that it will cease to hold when
the disk loses contact with the surface, \ie, when $F_z = 0$, at which time
$\ddot\alpha = - g/a$, or equivalently $d^2(1/\Omega^2)/dt^2 = - k$. 
Taking the derivative of eq.~(\ref{e107}), the maximum angular velocity is
\begin{equation}
\Omega_{\rm max} = \left( {k (\beta + 2)^2  \over 2 \beta}
\right)^{1 / 2 (\beta + 1) } C^{1 / (\beta + 1)},
\label{e108}
\end{equation}
which occurs at time $t_{\rm max}$ given by 
\begin{equation}
t_0 - t_{\rm max} = \left( {2 \beta \over k (\beta + 2)^2 } 
\right)^{(\beta + 2) / 2 (\beta + 1)}  C^{-1 / (\beta + 1)} .
\label{e109}
\end{equation}
%At that time,
%\begin{equation}
%\alpha_{\rm min}
% = \left( {3 \over 5 \epsilon} \right)^{2/3}
% \left( {g \over ak} \right)^{1/3} 
%\left( {2 \over 9} \right)^{1/2}
%\left( {3 \over 5 \epsilon} \right)^{1/3}
% \left( {g \over ak} \right)^{1/6} \left( {g \over a} \right)^{-1/2}
% = \left({2 \epsilon^2 (g/a)^{(\beta - 1)} \over 25  k^\beta} 
%\right)^{1/(\beta + 1)}.
%\label{e110}
%\end{equation}
%For a uniform disk with $k = 1/4$, and for the model of friction with
%$\beta = 1$, this gives $\alpha_{\rm min} = 0.57 \epsilon$.
%If the bump-spacing parameter $\epsilon$
%had a value of 0.1, then $\alpha_{\rm min} \approx 3.4^\circ$, 
%which is roughly as observed.

In Moffatt's model based on viscous
drag of the air between the disc and the surface \cite{Moffatt}, $\beta = 4$,
\begin{equation}
\alpha = \left( {2 \pi \eta a \over m} (t_0 - t) \right)^{1/3},
\label{e111}
\end{equation}
where $\eta = 1.8 \times 10^{-4}$ g-cm$^{-1}$-s is the viscosity of air,
and
\begin{equation}
\Omega = \sqrt{g \over ak} \left( {m / 2 \pi \eta a \over t_0 - t} 
\right)^{1/6}.
\label{e112}
\end{equation}
This model is notable for having no free parameters.
% This also yields $\alpha_{\rm min}$ of a few degrees, and hence a similar 
%value for $\Omega_{\rm max}$.  
%Formally, the air-drag model is the same as a 
%rolling-friction model with $\beta = 4$.

The main distinguishing feature between the various models for friction is the
different time dependences (\ref{e107}) for the angular 
velocity $\Omega$ as angle $\alpha$ decreases.  

A recent report \cite{Engh} indicates that the total times of spin of coins in
vacuum and in air are similar, which suggests that air drag is not the dominant
mechanism of energy dissipation.  Such results do not preclude that air drag
could be important for disks of better surface quality, and hence lower rolling
friction, or that air drag becomes important only during the high-frequency
motion as time $t$ approaches $t_0$.

\begin{figure}[htp]  % h = here, t = top, b = bottom, p = new page
\begin{center}
\vspace{0.1in}
\includegraphics[width=3.0in]{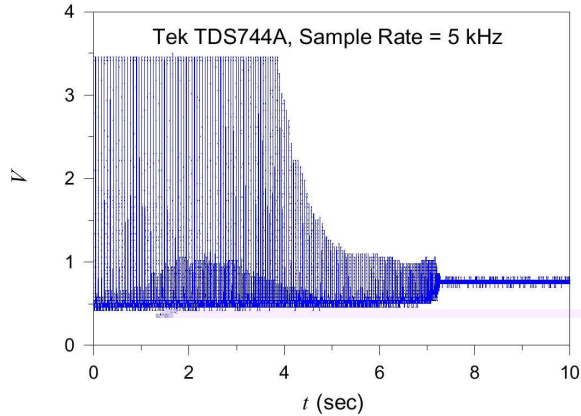}
\parbox{5.5in} % change 5.5in to \hsize for full-width caption
{\caption[ Short caption for table of contents ]
{\label{tek04} A 10-s record at 5 kS/s of the spinning of a Tangent Toy Euler's
Disk
\cite{Tangent} as observed by a phototransistor %\cite{Taos} 
that detected light reflected off the disk. 
}}
\end{center}
\end{figure}

\begin{figure}[htp]  % h = here, t = top, b = bottom, p = new page
\begin{center}
\vspace{0.1in}
\includegraphics[width=3.0in]{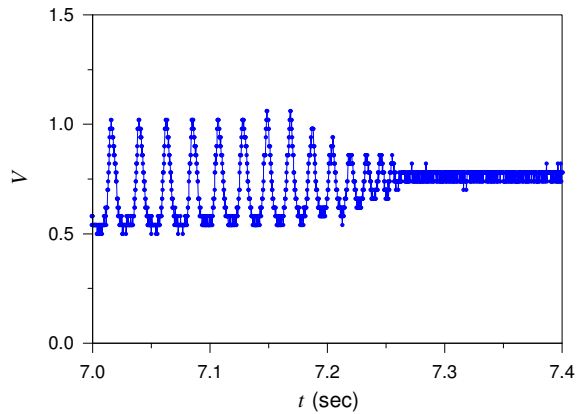}
\parbox{5.5in} % change 5.5in to \hsize for full-width caption
{\caption[ Short caption for table of contents ]
{\label{tek01} The last 0.25 s of the history of the spinning disk shown
in Fig.~\ref{tek04}. 
}}
\end{center}
\end{figure}

\begin{figure}[htp]  % h = here, t = top, b = bottom, p = new page
\begin{center}
\vspace{0.1in}
\includegraphics[width=3.0in]{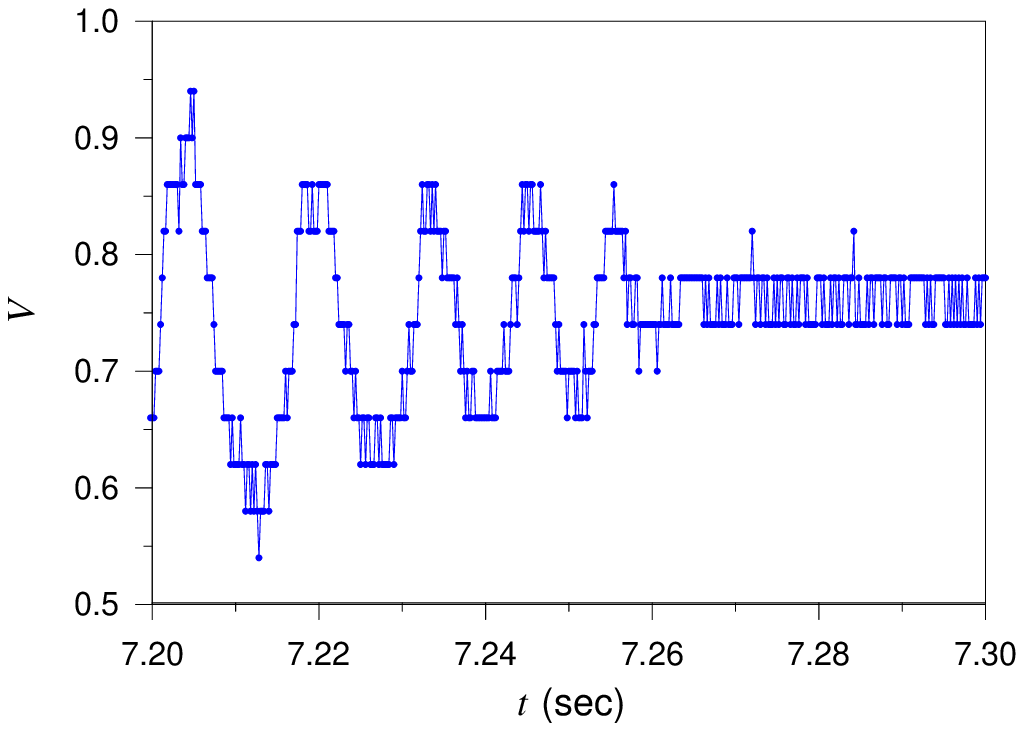}
\parbox{5.5in} % change 5.5in to \hsize for full-width caption
{\caption[ Short caption for table of contents ]
{\label{tek02} The last 0.06 s of the history of the spinning disk shown
in Figs.~\ref{tek04} and \ref{tek01}. 
}}
\end{center}
\end{figure}

To help determine whether any of the above models corresponds to the
practical physics, 
we have performed an experiment using a Tangent Toy Euler's Disk 
\cite{Tangent}.  The spinning disk was illuminated by a flashlight whose
beam was reflected off the surface of the disk onto a phototransistor
\cite{Taos} whose output was recorded by a digital oscilloscope \cite{Tek}
at 5 kS/s.
The record length of 50,000 samples permitted the last 10 seconds of the
spin history of the disk to be recorded, as shown in 
Figs.~\ref{tek04}-\ref{tek02}.

\begin{figure}[htp]  % h = here, t = top, b = bottom, p = new page
\begin{center}
\vspace{0.1in}
\includegraphics[width=3.0in]{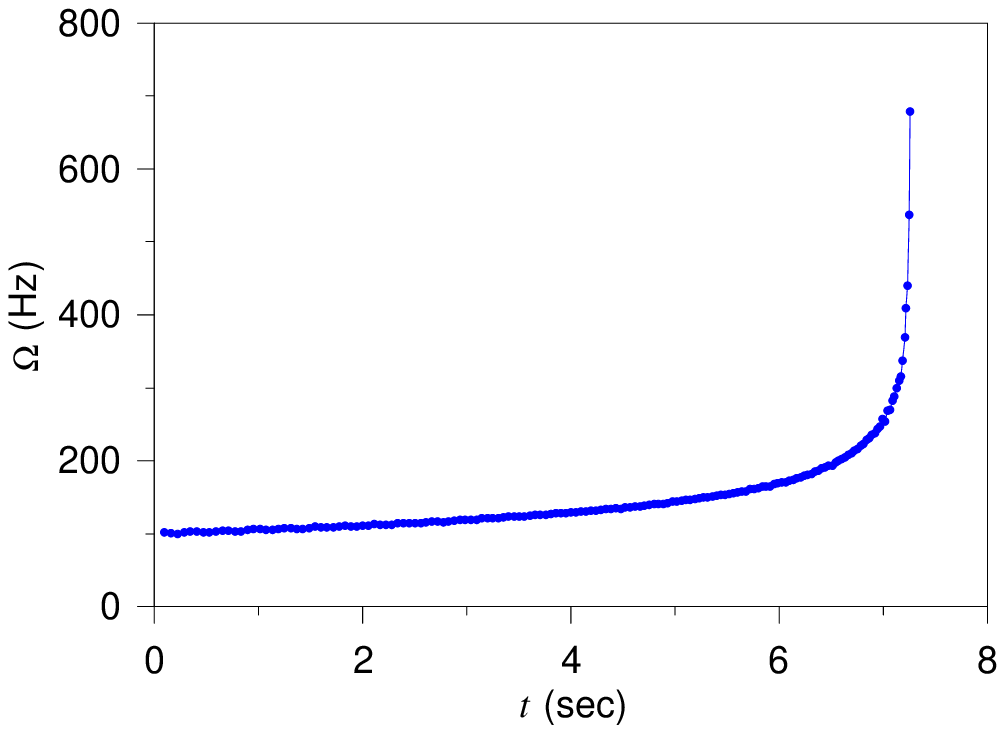}
\parbox{5.5in} % change 5.5in to \hsize for full-width caption
{\caption[ Short caption for table of contents ]
{\label{Omega} $\Omega$ \vs\ $t$ deduced from the data shown in 
Figs.~\ref{tek04}-\ref{tek02}. 
}}
\end{center}
\end{figure}

The analysis of the data shown in Figs.~\ref{tek04}-\ref{tek02} consisted of
identifying
 the time $t_i$ of the peak of cycle $i$  of oscillation as the mean of
the times of the rising and falling edges of the waveform at roughly one half
the peak height.  The average angular frequency for each cycle was calculated
as $\Omega_i = 2 \pi / (t_{i+1} - t_i)$, as shown in Fig.~\ref{Omega},
and the rate of change of angular
frequency was calculated as $d\Omega_i /dt = 2(\Omega_{i+1} - \Omega_i) /
(t_{i+2} - t_i)$.  The angular frequency of the last analyzable cycle was
$\Omega_{\rm max} = 680$ Hz.

\begin{figure}[htp]  % h = here, t = top, b = bottom, p = new page
\begin{center}
\vspace{0.1in}
\includegraphics[width=3.0in]{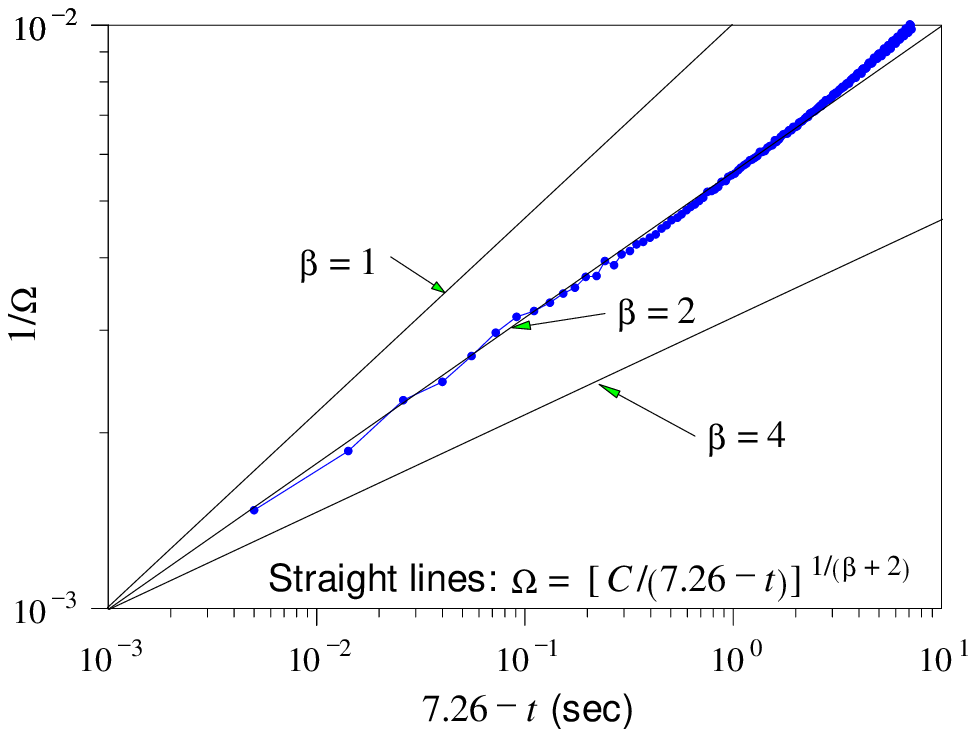}
\parbox{5.5in} % change 5.5in to \hsize for full-width caption
{\caption[ Short caption for table of contents ]
{\label{tek726} $1/\Omega$ \vs\ $t_0 - t$ for $t_0 = 7.26$ s, using
the data shown in Figs.~\ref{tek04}-\ref{tek02}.  The straight lines
illustrate the behavior expected according to eq.~(\ref{e107}) for various 
values of parameter $\beta$.
}}
\end{center}
\end{figure}

\begin{figure}[htp]  % h = here, t = top, b = bottom, p = new page
\begin{center}
\vspace{0.1in}
\includegraphics[width=3.0in]{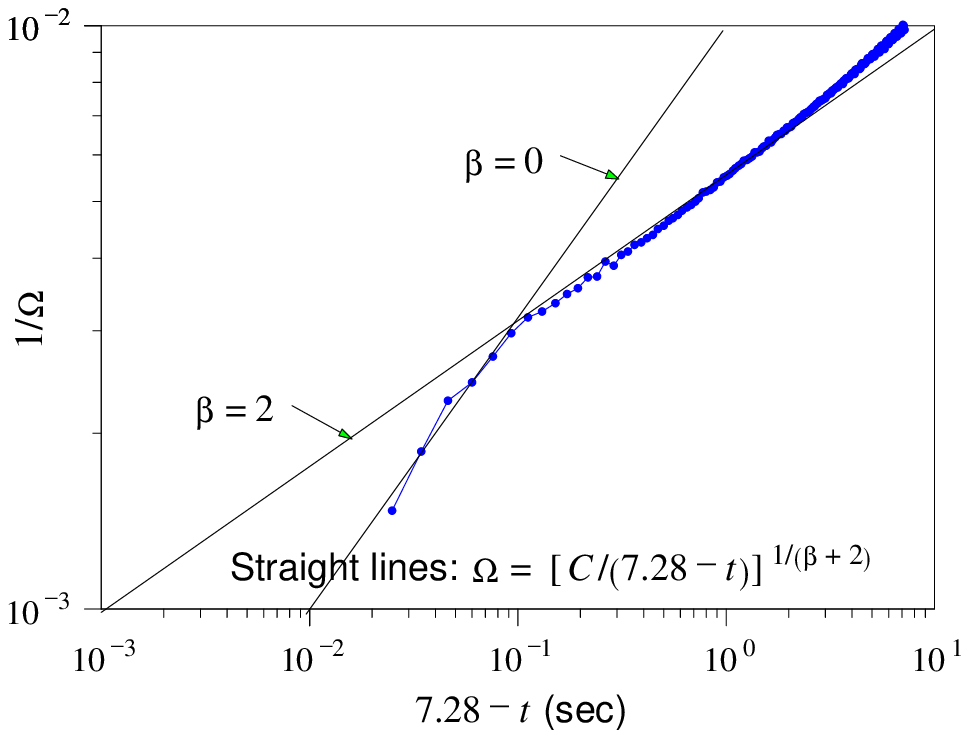}
\parbox{5.5in} % change 5.5in to \hsize for full-width caption
{\caption[ Short caption for table of contents ]
{\label{tek728} $1/\Omega$ \vs\ $t_0 - t$ for $t_0 = 7.28$ s, using
the data shown in Figs.~\ref{tek04}-\ref{tek02}. 
}}
\end{center}
\end{figure}

The data can be conveniently compared to the result (\ref{e107}) in the form
\begin{equation}
{1 \over \Omega} = \left( {t_0 - t \over C} \right)^{1/(\beta + 2)}
\label{e120}
\end{equation}
 via a
log-log plot of $1/\Omega$ \vs\ $t_0 - t$, given an hypothesis as to $t_0.$
Inspection of Fig.~\ref{tek02} suggests that $t_0$ is in the range 7.26-7.28 s
for our data sample.  Figures~\ref{tek726} and \ref{tek728} show plots of
$1/\Omega$ \vs\ $t_0 - t$ for $t_0 = 7.26$ and 7.28 s, respectively.  The
straight lines are not fits to the data, but illustrate the behavior expected
according to eq.~(\ref{e107}) for various values of parameter $\beta$.

A larger value of $t_0$ has the effect of lowering the apparent value of
parameter $\beta$ for the last few cycles of the motion.  Figure~\ref{tek02}
suggests that $t_0$ could hardly be less than $7.26$ s, for which case a value
of $\beta = 2$ would fit the entire data sample rather well.  
%This value of $\beta$ corresponds to friction that varies directly with the
%velocity of the point of contact.

In view of the uncertainty in assigning a value to the time $t_0$, it is
interesting to ask at what time $t$ does the time remaining equal
exactly one period of the motion, \ie, when does $t_0 - t = 2 \pi / \Omega(t)$?
For $\beta = 2$, the answer from eq.~(\ref{e107}) is when
$\Omega = (C / 2 \pi)^{1/3}$.  From Fig.~\ref{tek726} we estimate that
$C^{-1/4} = 0.0055$, and hence $\Omega \approx 560$ Hz when the remaining time
of the
motion is $2 \pi / \Omega = 0.011$ s.  Recall that the last cycle analyzable 
in our data sample yielded a value of 680 Hz for $\Omega$.  The preceding
 analysis tells us that time $t_0$
cannot be more than about 0.01 s after the last observable peak in the data,
which suggests that $t_0$ is closer to 7.26 than to 7.28 s, and that 
Fig.~\ref{tek726} is the relevant representation of the experiment.

For $\beta = 2$, the spinning disk is predicted by eq.~(\ref{e109}) to lose
contact with the horizontal surface when $t_0 - t = C^{-1/3} = 0.001$ s for
$C^{-1/4} = 0.0055$.  The instantaneous angular frequency at that time is
predicted by eq.~(\ref{e108}) to be $\Omega = C^{1/3} = 1030$ Hz.  These
values are, of course, beyond those for the last analyzable cycle in the
data.

\begin{figure}[htp]  % h = here, t = top, b = bottom, p = new page
\begin{center}
\vspace{0.1in}
\includegraphics[width=3.0in]{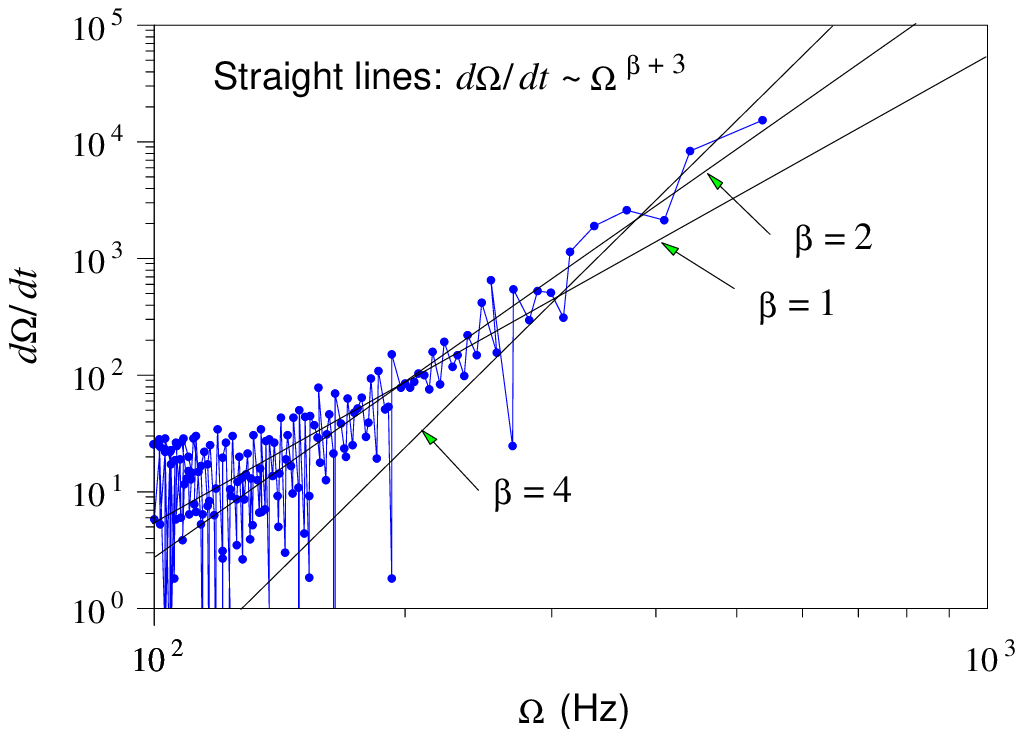}
\parbox{5.5in} % change 5.5in to \hsize for full-width caption
{\caption[ Short caption for table of contents ]
{\label{Odot} $d\Omega/dt$ \vs\ $\Omega$ for the data shown in 
Figs.~\ref{tek04}-\ref{tek02}.  The straight lines
illustrate the behavior expected according to eq.~(\ref{e121}) for various 
values of parameter $\beta$.
}}
\end{center}
\end{figure}

The question as to the value of $t_0$ can be avoided by noting 
\cite{Chatterjee} that the
time derivative of eq.~(\ref{e107}) yields the relation
\begin{equation}
{d \Omega \over dt} \propto \Omega^{\beta + 3}.
\label{e121}
\end{equation}
However, $d\Omega / dt$ must be calculated from differences of differences of 
the times of the peaks in the data, so is subject to greater uncertainty than
is $\Omega$.  Figure~\ref{Odot} shows a log-log plot of $d\Omega/dt$ \vs\
$\Omega$ together with straight lines illustrating the expected behavior for
various values of $\beta$.  Again, $\beta = 2$ is a consistent description of
the entire data sample.
% but the fluctuations during the last few cycles do not
%exclude the possibility that the value of $\beta$ changed somewhat then.
The value of $\beta = 0$ suggested by Fig.~\ref{tek728} based on $t_0 = 7.28$
s is quite inconsistent with Fig.~\ref{Odot}. 

The results of our experiment on the time history of the motion of a spinning
disk are not definitive, but are rather consistent with the dissipated
power being proportional to the square of the velocity
of the point of contact, and hence the dissipative force varying linearly
with velocity.  Our experiment cannot determine whether during the 0.01 s 
beyond the last full cycle of the motion
 an additional dissipative mechanism such as air friction with
power loss
proportional to the fourth power of the velocity \cite{Moffatt} became
important.

\section{Summary of the Motion of a Disk Spun Initially About a Vertical 
Diameter}

If a uniform disk 
 is given a large initial angular velocity about a vertical diameter,
and the initial horizontal velocity of the center of mass is very small, 
the disk will
``sleep'' until friction at the point of contact reduces the angular
velocity below that of condition (\ref{e407}) (secs.~8).  
The disk will then appear to
fall over rather quickly into a rocking motion with angle
$\alpha < 90^\circ$ (sec.~9).  After this, the vertical angular velocity $\Omega$
will increase ever more rapidly, while angle $\alpha$ decreases (sec.~5), 
until the disk
loses contact with the table at a value of $\alpha$ of a few degrees 
(sec.~11).
The disk then quickly settles on to the horizontal surface.  One hears sound
at frequency $\Omega / 2 \pi$, which becomes dramatically higher until the
sound abruptly ceases.  But if one observes a figure on the face of the
disk, this rotates every more slowly and seems almost to have stopped moving
before the sounds ceases (sec.~4).

If the initial motion of the disk included a nonzero initial velocity in
addition to the spin about a vertical diameter, the center of mass will
initially move in a circle (sec.~4).  If the initial
vertical angular velocity is small, the disc will roll in a large circle,
tilting slightly inwards until the rolling angular velocity $\omega_1$
drops below that of condition (\ref{e49}).  
While in most cases
the angle $\alpha$ of the disk will then quickly drop to $60^\circ$ or so
(sec.~6),
occasionally $\alpha$ will rise back towards $90^\circ$ before falling
(sec.~9).
As the disk rolls and spins, the center of mass traces an inward spiral on
average (sec.~5), but nutations about this spiral can be seen, 
often accompanied by a rattling sound (sec.~6). 
The nutation is especially prominent for 
$\alpha \approx 10-15^\circ$ at which time a very low beat frequency between 
that of primary spin and that of the small oscillation can be observed
(sec.~6).  As $\alpha$ 
decreases below this, the radius of the circle traced by the center of mass
becomes very small, and the subsequent motion is that of a disk without
horizontal center of mass motion (secs.~4 and 8).

\medskip

The authors thank A.~Chatterjee, C.G.~Gray, H.K.~Moffatt,
M.G. Olsson and A.~Ruina for 
insightful correspondence on this topic.

\end{document}